\documentstyle[preprint,eqsecnum,aps]{revtex}
\begin{document}
\draft
\preprint{BU-HEP-94-23}

\title{SUBTLETIES AND FANCIES IN \\
        GAUGE THEORY NON TRIVIAL VACUUM \footnotemark[1]}

\footnotetext[1]{This work is supported in part by funds
provided by the U. S. Department of Energy (D.O.E.) under
contracts \#DE-FG02-91ER40676}

\author{A.~R.~Levi}
\address{Department of Physics, Boston University\\
         590 Commonwealth Avenue, Boston, MA 02215}

\date{June 6, 1994}
\maketitle
\begin{abstract}
The one loop effective potential for a non-Abelian gauge
configuration is analyzed using the background field method.
The Savvidy result and the non-Abelian ansatz,
the other alternative possible background that
generates a constant color magnetic field configuration,
are compared.
This second possibility is very interesting because
it avoids the possible
coordinate singularity, ${\rm Det}B_i^a=0$, and
it is easy to implement in lattice simulations.
We emphasize the interesting dependence of the potential
by the gauge fixing parameter $\alpha$, when
the loop expansion is performed around a non trivial
background configuration.
Finally, we point out some crucial
differences in analyzing the vacuum structure
between non-Abelian gauge theories and  the cases of scalar and
Abelian gauge theories.

\

{\it To appear in Proceeding of the ``Workshop on Quantum
Infrared Physics'', Paris, June 1994.}
\end{abstract}
\widetext

\section{INTRODUCTION}

Even after many years of intensive efforts, the vacuum
structure of non-Abelian gauge theories still remains
elusive to us.
For a recent article where several aspects of the non-trivial
vacuum for these theories have been analyzed
see \cite{noi}, and for a comprehensive revue see \cite{muller}.

As a prelude to truly non-perturbative evaluation of the
effective I.R. structure of non-Abelian theories in terms of
lattice,
the conventional approach is to analyze the
effective action using the so-called background
field method.

For scalar and Abelian gauge theories, we can assert that we understood
very well their I.R. structures.
In these cases, the vacuum is not ill-defined and this allows the
construction of the complete physical space of the theory by acting with
creator operators on the vacuum state. This procedure it is not
ambiguous and it is formally well understood.

However, the situation of the non-Abelian
case is very different, and very little has to do
with the Abelian and scalar cases.
For non-Abelian theories, there is no simple
way to decompose the field into non-interacting ``field oscillators'';
they are non-linear and coupled together.
To be more precise, there are two different
regions: for large Euclidean momenta the effective coupling
constant $g^2(p)$ is relatively small,
and the perturbative picture of the plane
gluonic waves is reasonable.
Many perturbative results can be achieved in this region.
For example, the reason why deep-inelastic scattering gives results
which are so clear is not only
because in high momenta transfer the gluonic gauge
coupling is small, but also
because the vacuum is so far away from the
measured states with high momenta that the
experimental data are independent of the vacuum
structure.

The real problem is only with the momenta of the
order of 1~GeV or smaller, where the infrared
structure of the field theory shows up.
For small momenta, the coupling is strong and
it is not reasonable to assume that the
non-Abelian gauge fields are distributed in
space-time as plane waves.
This suggests a classification of the vacuum models.
The first model assume that the non-perturbative field is concentrated
in some space-time region, which is the instanton-type vacuum.
If it is concentrated in space, but on a line along time direction,
is called a soliton-type vacuum.
If it is concentrated on a two-dimensional surface, it is
called a string-type vacuum.
However, there is also the possibility that
there is a homogeneous-type vacuum that will be referred to as the
``non-trivial vacuum'' which we
will discuss here.

The SU(2) case should not be fundamentally different from other
non-Abelian gauge theory. In fact, up to now, there is
no reason why the QCD vacuum should not be in the same
universality class of the SU(2) case.
On the other hand, SU(2) allows
explicit calculations that
are still inaccessible for the SU(3) case.

Beside lattice computations, the only available
method to investigate the properties of Yang-Mills theories vacuum is
computation of the effective potential
in the presence of a background gauge \cite{abbott}.
This manifest gauge invariant scheme is based on the observation that
the loop expansion corresponds to
an expansion in the parameter $\hbar$ which multiplies
the entire action. Hence, a shift of the fields, or
a redefinition of the division of the Lagrangian
into free and interacting parts can be performed
at any finite order of the loop expansion without violating the manifest
gauge invariance.

Analogous conclusions can be obtained by
studying the beta function. This is not surprising
because the beta function is dependent
on the ultraviolet part of the effective potential.

Further note that
this procedure is straightforward only if the effective potential
does not contain operators which are irrelevant at the ultraviolet
fixed point but become relevant in the infrared.
In this case, the $\beta$-function by construction,
depends only on the ultraviolet relevant coupling constants and knowledge
of it is not sufficient
to investigate the infrared properties of the effective potential.

In doing the one-loop calculation, we encountered several
technical points, which were known in the literature but not
emphasized enough. We specially
want to stress the gauge fixing $\alpha$
parameter dependence. Therefore, we decided to briefly go through
the standard background field method in gauge
theories with those technical points in mind.

\section{ EFFECTIVE POTENTIAL FORMALISM}

Before we launch an extensive numerical simulation,
it is still worthwhile to calculate the one-loop
effective potential for the non-Abelian background
field Eq.(\ref{nonabel}).
Even though it is not totally trustworthy, the loop
expansion can still provide indicative information.
Furthermore, to examine whether the qualitative
feature of the one-loop effective
potential strongly depends on the choice of the background
field is also interesting. In particular, we would like to
find out whether the coordinate singularity has any connection
with the existence of the imaginary part in the effective
potential. We will prove that the imaginary part cannot
be eliminated by avoiding the coordinate singularity of
${\rm Det}B_i^a=0$.

We will now outline the basic steps in the background
field method in gauge theories \cite{abbott}. There are two
reasons to go through the well-known method. The first is to
establish our own notations. The second is to emphasize two
technical aspects; the question of whether it is necessary to
require the background field to satisfy the classical equation of
motion and the gauge choice in the evaluation of the functional
integral. When possible we follow the convention of Abbott
\cite{abbott}.

The generating functional in the background field method
(in Euclidean space) is defined throughout
\begin{equation}
    {\tilde Z}[J,A]=\int[dQ]\,
    \det\biggl[{\delta G^a\over\delta\omega^b}\biggr]
    \exp\biggl\{-\int d^dx\bigl[{\cal L}(A+Q)
    +J_\mu^a Q_\mu^a\bigr]\biggr\}
    \prod_{x,a} \delta[G^a],
\label{zstart}
\end{equation}
where $G^a\equiv\partial_\mu Q_\mu^a+g f^{abc} A_\mu^b Q_\mu^c
=D_\mu^{ab}(A)Q_\mu^b$ is the background field gauge condition
and $\det\left[\delta G^a/\delta\omega^b\right]$ is the
corresponding Jacobian. It is clear from the above definition
the background field $A_\mu^a$ is fixed and should be regarded
as an external parameter in the process of doing the $Q$--integral.
Note also that we do not exponentiate the gauge constraint
$\prod_{x,a} \delta[G^a]$, in contrast with what was usually done.
The necessity of enforcing the gauge constraint with explicit
delta-functions in Eq.(\ref{zstart}) will be discussed
in detail.

It is easy to verify that ${\tilde Z}[J,A]$ is invariant under
the infinitesimal gauge transformations
\begin{mathletters}
\label{ginv}
\begin{equation}
      \delta A_\mu^a=-f^{abc}\omega^b A_\mu^c
      +{1\over g}\partial_\mu \omega^a \, ;
\end{equation}
\begin{equation}
      \delta J_\mu^a=-f^{abc}\omega^b J_\mu^c \, .
\end{equation}
\end{mathletters}
As a consequence of this invariance, the effective action
in the background field method,
defined as the Legendre transform of $\tilde Z$:
\begin{equation}
    {\tilde \Gamma}[{\tilde Q},A]=-\ln {\tilde Z}[J,A]
    +\int d^dx\,J_\mu^a {\tilde Q}_\mu^a \, ,
\label{lt}
\end{equation}
with ${\tilde Q}_\mu^a=\delta \ln {\tilde Z}/ \delta J_\mu^a$,
is invariant under
\begin{mathletters}
\label{gt}
\begin{equation}
      \delta A_\mu^a=-f^{abc}\omega^b A_\mu^c
      +{1\over g}\partial_\mu \omega^a \, ;
\end{equation}
\begin{equation}
       \delta {\tilde Q}_\mu^a=-f^{abc}\omega^b {\tilde Q}_\mu^c \, .
\end{equation}
\end{mathletters}
As shown by Abbott \cite{abbott}, when ${\tilde Q}_\mu^a=0$,
${\tilde \Gamma}[0,A]$, it coincides with the usual effective action.
Since Eq.(\ref{gt}) is a pure gauge transformation when $\tilde Q$
vanishes, ${\tilde \Gamma}[0,A]$ must be a gauge invariant
functional of $A_\mu^a$.

A standard way to evaluate ${\tilde Z}[J,A]$ explicitly is
to make a loop expansion. For the purpose of convenience, we
exponentiate the Jacobian $\det[\delta G^a/ \delta\omega^b]$
and gauge constraints $\prod_{x,a} \delta[G^a]$ by introducing
the Faddeev-Popov ghost fields ($\theta^a$ and $\bar\theta^a$)
and a real scalar auxiliary field $\sigma^a$, respectively,
\begin{eqnarray}
     \tilde{Z}[J,A] &=&\int[dQ][d\theta][d\bar\theta][d\sigma]\,
     \exp\biggl\{-\int d^dx\bigl[{\cal L}(A+Q)
     +{\bar\theta^a}
     D_\mu^{ac}(A)D_\mu^{cb}(A)\theta^b + \nonumber \\
     & &+2i\sigma^a D_\mu^{ab}(A)Q_\mu^b
     +J_\mu^a Q_\mu^a\bigr]\biggr\} \, .
\label{zend}
\end{eqnarray}
If we only want to calculate $\tilde Z[J,A]$ to one-loop order,
then it is equivalent to evaluating Eq.(\ref{zend})
in the steepest descent approximation,
\begin{eqnarray}
\tilde{Z}[J,A]&\approx& {\tilde Z}_1[J,A]\equiv
    \int[dQ][d\theta] [d\bar\theta][d\sigma]
    \exp\biggl\{-\int d^dx\bigl[{\cal L}(A)
    +{\cal L}^{(1)}(A,Q) + \nonumber \\
    & & {\cal L}^{(2)}(A,Q)
    +{\bar\theta^a} D_\mu^{ac}(A)D_\mu^{cb}(A)\theta^b
    +2i\sigma^a D_\mu^{ab}(A)Q_\mu^b
    +J_\mu^a Q_\mu^a\bigr]\biggr\} \, ,
\label{z1}
\end{eqnarray}
where
\begin{equation}
     {\cal L}^{(1)}(A,Q)=-Q_\mu^a D_\mu^{ab}(A)F_{\mu\nu}^b(A)\, ;
\end{equation}
\begin{equation}
     {\cal L}^{(2)}(A,Q)=Q_\mu^a\biggl[
     -{1\over 2}\left(D_\rho(A)D_\rho(A)\right)^{ab}\delta_{\mu\nu}
     +ig F_{\mu\nu}^{ab}(A)\biggr]Q_\nu^b
     \equiv Q_\mu^a M_{\mu\nu}^{ab}(A) Q_\nu^b \, .
\label{ma}
\end{equation}
In arriving at the expression of ${\cal L}^{(2)}(A,Q)$, the gauge
constraint condition $D^{ab}_\mu Q_\mu^b=0$ has been used.

We can now do the $Q$--integral in ${\tilde Z}_1[A,J]$ and
make the Legendre transform as in Eq.(\ref{lt}). By setting
${\tilde Q}_\mu^a=0$ the final expression for the effective
action to one-loop order can be written as
$\Gamma[A]=-\ln {\tilde Z}_1[A]$, with
\begin{eqnarray}
     {\tilde Z}_1[A]&=&e^{-\int d^dx {\cal L}(A)}\cdot
     \det\bigl[-D(A)D(A)\bigr] \cdot \nonumber \\
     & & \int [dQ][d\sigma]
     \exp\biggl\{-\int d^dx\bigl[Q_\mu^a M_{\mu\nu}^{ab}
     (A)Q_\nu^b + 2i\sigma^a
     D_\mu^{ab}(A)Q_\mu^b\bigr]\biggr\}  \nonumber \\
     &=&e^{-\int d^dx {\cal L}(A)}\cdot
     \det\bigl[-D(A)D(A)\bigr] \cdot
     \biggl\{\det M_{\mu\nu}^{ab}\cdot\det N^{ab}\biggr\}^{-1/2} \, ,
\label{aeff}
\end{eqnarray}
with $N^{ab}=-D_\mu^{ac} (M^{-1})^{ce}_{\mu\nu}D_\nu^{eb}$.
The first factor is the classical contribution. The second and
third factors are the one-loop quantum corrections due to the
ghost field ($\theta$) and fluctuation field ($Q$), respectively.
Notice that the linear term ${\cal L}^{(1)}(A,Q)$ drops out
automatically in the calculation process.
To remove the linear term there is no need to
require that the background field satisfy the classical equation
of motion $D_\mu^{ab}(A)F_{\mu\nu}^b(A)=0$. This independence of
the linear term remains true to all orders in loop expansion,
because the effective potential is a sum of all one-particle
irreducible diagrams with $A$ fields on the external lines and
$Q$ fields on internal lines (when ${\tilde Q}_\mu^a=0$). In
fact, the linear term is always compensated by the source $J_\mu^a$.
Remember that the physical limit in the background field method
is ${\tilde Q}_\mu^a=0$, not $J_\mu^a=0$. On the other hand, it
should be pointed out that to require the background field to
satisfy the equation of motion, in order to have
a good chance to represent the true minimum of the action,
is an entirely different matter.

Since the background field is a constant in space-time it
is convenient to work in momentum-space. Using the formula
$\det M=\exp({\rm Tr}\ln M)$ combined with Eq.(\ref{aeff}),
we have the expression for the effective potential to
one-loop order
\begin{equation}
  V_{\rm eff}(h)={\cal V}_{classical}
  -\int{d^d p\over(2\pi)^d}\ln G(h;p)+{1\over2}\int{d^d p\over(2\pi)^d}
  \biggl\{\ln M(h;p)+\ln N(h;p) \biggr\} \, .
\label{veff2}
\end{equation}

An eventual non-positive definiteness of $M(h;p)$ and $N(h;p)$ for small
$p^2$ indicates that the background field
is only a saddle point in configuration space and therefore
is not stable.

Eq. (\ref{veff2}) must be regularize to
eliminate divergent constants. The standard method is to use
the regularization of Salam and Strathdee \cite{salam},
which is a variation of Schwinger's proper time method
\cite{schwinger}.
Peculiar tricks have to be implemented depending
in how many dimensions the calculation is done.
For example, because the three dimensional
Yang-Mills theory is super renormalizable, the only divergence
we encounter is an overall additive constant. The regularization
procedure includes three steps. First, an integral
representation for logarithmic function is used. For real $E$,
positive or negative,
\begin{equation}
    \ln(E-i\delta)={1\over\epsilon}
    -{i^\epsilon\over\epsilon\Gamma(\epsilon)}
     \int dt\, t^{\epsilon-1}\, e^{-it(E-i\delta)} \, ,
\label{ssr2}
\end{equation}
in the limit of $\epsilon\rightarrow 0^+$. The $i\delta$
in Eq.(\ref{ssr2}) with $\delta\rightarrow 0^+$ is to ensure
the convergence of the integral. When $E$ is explicitly
complex, Eq.(\ref{ssr2}) can be easily generalized. Then the
momentum integration can be done using
\begin{equation}
         \int {d^3p\over(2\pi)^3}e^{-itp^2}={1\over 8\pi^{3/2}}
         (it)^{-3/2} \, .
\end{equation}
The remaining $t$--integral can be converted into a Gamma
function through the contour integral technique. Finally,
the limit of $\epsilon\rightarrow 0$ is taken.

At this point we would like to comment on the implementation of
the gauge condition $\prod_{x,a}\delta[G^a]$ in Eq.(\ref{zstart}).
The standard approach is to exponentiate this factor,
\begin{equation}
     {\tilde Z}[J,A;\alpha]=\int[dQ]\,
     \det\biggl[{\delta G^a\over\delta\omega^b}\biggr]
     \exp\biggl\{-\int d^dx\bigl[{\cal L}(A+Q)
     -{1\over 2\alpha}(G^a)^2+J_\mu^a Q_\mu^a\bigr]\biggr\} \, ,
\label{zexp}
\end{equation}
where $\alpha$ is the gauge fixing parameter. In order to write
Eq.(\ref{zstart}) into the form of Eq.(\ref{zexp}), one has to
generalize the gauge condition as
$\prod_{x,a}\delta[G^a-f^a]$, with $f^a$ being arbitrary,
and then to integrate out $f^a$ with the weighting factor
$\exp[-(f^a)^2/2\alpha]$. If ${\tilde Z}[J,A]$ with the
generalized gauge condition were independent of $f^a$, as in
the case of zero background field $A_\mu^a=0$,
${\tilde Z}[J,A;\alpha]$ would be independent of the gauge
parameter $\alpha$ and therefore the exact equivalence between
Eq.(\ref{zstart}) and Eq.(\ref{zexp}). However, in the presence
of non-vanishing background field $A_\mu^a$, ${\tilde Z}[J,A]$
in general would depend on $f^a$, as can be easily seen in
Eq.(\ref{zend}) by adding a term $2i\sigma^a f^a$ to the
exponential. Physically, this dependence on $f^a$ is natural,
because there would exist correlations between $f^a$
and certain combinations of the background field, such as
$(D_\mu^{ab}(A)f^b)^2$ when $A_\mu^a$ is non-vanishing.
If indeed ${\tilde Z}[J,A]$ depended on $f^a$, then the integral
\begin{equation}
   {\tilde Z}[J,A;\alpha]\equiv
   \int [df] \,{\tilde Z}[J,A] \bigg|_{G^a=f^a}\,
   \cdot\exp\biggl\{-{1\over 2\alpha}(f^a)^2\biggr\}
\end{equation}
would have to depend on $\alpha$ in general, which in turn implies
that Eq.(\ref{zexp}) could not be equivalent to Eq.(\ref{zstart})
at all times.

This gauge dependence of the effective action calculation
has long been recognized in the literature \cite{vilkovisky}.
It was argued by Vilkovisky that the correct choice of the
gauge parameter is to take $\alpha\rightarrow 0$ limit, or to
choose the Landau background field gauge.
Since $\exp[-x^2/2\alpha]\propto\delta(x)$ in the limit of
$\alpha=0$, Eq.(\ref{zexp}) is formally equivalent to
Eq.(\ref{zstart}) in the Landau background field gauge. It
should be emphasized that Eq.(\ref{zexp}) would still define
a gauge invariant effective action with respect to the
background field $A_\mu^a$. It is only the functional
form of the effective action that depends on the gauge
parameter when $\alpha\ne 0$.
Since the $\beta$--function is gauge
(or $\alpha$) independent in the dimensional regularization
and minimal subtraction scheme \cite{gross}, even though the finite
part is explicitly $\alpha$ dependent,
this has generated much confusion.
Finally, note that in the context of finite
temperature QCD, this gauge dependence has been emphasized by
Hansson and Zahed \cite{hansson}.

\section{Savvidy vs non-Abelian}

Because the effective potential is the truncation
of the effective action at zero momenta, we are
seeking solutions that give a constant magnetic field
configurations.
There are only two possibilities called, respectively, the
Savvidy (or Abelian), and the non-Abelian ansatz.
It has been proved \cite{brown} that
there are no other ways to generate a constant
magnetic field.

We will first analyze the Savvidy's case.
As in Savvidy's and Matinyan's pioneer work \cite{savvidy},
where they
calculated up to one-loop order the effective potential
for an Abelian background gauge field:
\begin{equation}
       A^a_\mu(x) = {1\over2} H \delta^{a3}
       (x_1 \delta_{\mu 2}  - x_2 \delta_{\mu 1}  )\, ,
\label{abel}
\end{equation}
which generates a constant color magnetic field
$B_i^a=H\delta_{i3}\delta_{a3}$
in the SU(2) Yang-Mills theory. It was then found,
remarkably, that the vacuum with this background field
is energetically favored over the perturbative vacuum
($H=0$). Unfortunately, a more careful analysis
\cite{nielsen} soon revealed that there exists an
imaginary part in the effective potential, indicating
that the background field Eq.(\ref{abel}) is not a
minimum but rather a saddle point in the configuration
space in the context of the loop expansion.
The subsequent work by the Copenhagen group
\cite{copenhagen}, still within the framework of
loop expansion, tried to remedy the physically
appealing picture of Savvidy's vacuum by introducing
inhomogeneity, leading to the so-called Copenhagen vacuum.

For the Savvidy ansatz,
it is easy to solve and find the eigenvalues of the
second derivative of the action,
because is  exactly the same operator of the
Landau levels problem, which has similar symmetry.
In fact, the corresponding
$M$ matrix is simply the Landau diamagnetic Hamiltonian.
Therefore, their eigenvalues are:
\begin{equation}
\label{lamarmora}
     \lambda(k)~=~ k^2 + (2n+1)~gH +2gH~S_z
\end{equation}
where $k=k_0^2 +k_z^2$, $n=0,1,2,...$ and $S_z=\pm 1$.
As expected from the symmetry of the problem, the modes
in the $x,y$ directions are quantized.
However, the behavior of the modes along the $z$ axis are labeled
by an arbitrary continuous parameter, $k_z$.
Moreover, note the difference from the Landau level where
the quantum number $S_z$ takes value $S_z=\pm 1/2$, which is the
electron polarization; whereas, in this case, we have the
polarization of the photon that give $S_z=\pm 1$, allowing the
possibility of negative eigenvalue when $(S_z=-1, n=0)$.

Using the technique discussed
in the previous section, it is possible to evaluate
the effective potential.
For the reader interested in an accurate discussion of the
eigenfunction and of their multiplicity see
\cite{maiani}.
For the Savvidy case, in (3+1) space-time dimensions the effective
potential up to one loop contributions is given by
\begin{equation}
\label{allende}
     V(H)~=~{1\over 2}H^2~+~{11\over 48\pi^2}~g^2H^2 \biggl(~
     \ln {gH\over \mu^2}~-~{1\over 2}~\biggr)~
     -i~{1 \over 8\pi}~g^2 H^2 ~+~\ldots \, .
\end{equation}

It is also interesting to look at the 3 dimensional case.
Because the theory in 3 dimensions is superrenormalizable,
we should better control the full procedure.
Moreover, Lattice simulations can be obviously performed easily,
both because of the volume, and because the scaling window is very
large due to the polynomial dependence of the renormalizable
quantity throughout the $\beta$ function, in contrast with the well
known exponential dependence of the 4 dimensions.
Therefore, several authors recently analyzed the 3 dimensional
case both analytically and with extensive lattice simulations
\cite{woloshyn}.
In 3 space-time dimensions, for the Savvidy ansatz, we have:
\begin{equation}
\label{allz}
     V(H)~=~{1\over 2}H^2~-~{1\over 2\pi^2}~ \biggl[
       1-{\sqrt{2} -1 \over 4\pi } \zeta
       \bigg( {3\over2}\biggr) \biggr]~(gH)^{3/2}~
       +i~ {1\over 12\pi}~(gH)^{3/2}~+~\ldots \, ,
\end{equation}
where $\zeta$ is the Riemann's Zeta-function.

The non-Abelian ansatz give a similar result in the structure,
but different results in the coefficients.
Moreover, the physical picture is not the same of Landau levels,
but is the same as a system with momenta coupled with spins.
The Non-Abelian ansatz is:
\begin{equation}
      A_0^a(x)=0,\,\,\,\,\, {\rm and}
      \,\,\,\,\, A_i^a(x)=h\delta_i^a \, ,
\label{nonabel}
\end{equation}
where $h$ is a constant in space-time.
This choice was first considered in a work of
Ambj\o rn, Nielsen and Olesen \cite{forg}.
In respect to the Savvidy, the non-Abelian has the crucial vantage
of being naturally implementable in lattice simulations.
Moreover, recently, in an unrelated but very interesting
work by Johnson and his collaborators \cite{johnson},
the Schr\"{o}dinger functional approach in terms of
magnetic field strength is applied to the SU(2)
Yang-Mills gauge theory.
In the explicit Schr\"{o}dinger functional
Hamiltonian they find that there is a factor
$1/{\rm Det}B_i^a$ in the kinetic energy term, similar
to the $1/r$ factor in the quantum mechanics
Schr\"{o}dinger equation in polar coordinate.
This factor may signal a potential coordinate
singularity for the color magnetic field in the
configuration space, similar to that of
the wavefunctions in quantum mechanics which have to
satisfy certain boundary conditions at $r=0$.
Since the Savvidy ansatz yields
${\rm Det}B_i^a=0$, due to its intrinsic Abelian nature,
it is desirable to seek an alternative background field,
which avoids this potential coordinate singularity.
The corresponding
color magnetic field for the non-Abelian ansatz is simply given by
$B_i^a=gh^2\delta_i^a$ and
obviously has ${\rm Det}B_i^a\ne 0$ when $h\ne 0$.
It is easy to recognize that the constant magnetic
field is generated by Eq.(\ref{nonabel}) through the
commutator terms in the field strength,
rather than the derivative terms as in the
Savvidy case.
Therefore, due to the non-trivial correlation
between its directions in space and color space,
the background field Eq.(\ref{nonabel})
evades the possible coordinate singularity.

In addition, since the non-Abelian background is
space-time independent, it is
much easier to put it on a lattice than the non-constant
background field Eq.(\ref{abel}). For example,
periodic boundary conditions are automatic.

For the non-Abelian case the second derivative of
the action in respect to the gauge fields
give the operator
\begin{equation}
    M^{ab}_{\mu \nu}~=~{1\over2}
    \bigg[ (p^2+2g^2h^2)\delta^{ab}\delta_{\mu\nu}
    +2gh ~(\sigma_{ab})^c p_c\delta_{\mu\nu}
    -2g^2h^2 ~(\sigma^{ab})_c (\sigma_{c\mu})_\nu  \biggr]\, ,
\label{mh_4}
\end{equation}
which is a $9\times9$ matrix for a given momentum $p$. This
matrix can be interpreted as the Hamiltonian of a relativistic
spin-1 ($\vec{\sigma}_s$) and color spin-1 ($\vec{\sigma}_c$)
boson, with the first term being the free particle part, the
second the $\vec{p}\cdot\vec{\sigma}_c$ and the third
$\vec{\sigma}_s\cdot\vec{\sigma}_c$. The last two terms can be
negative for low momentum, depending on the relative
orientations of $\vec{p}$,
$\vec{\sigma}_s$ and $\vec{\sigma}_c$.

For the non-Abelian ansatz,
in (3+1) dimensions the effective potential is
\begin{equation}
\label{chilosa}
     V(H)~=~{1\over 2}H^2~+~{21\over 48\pi^2}g^2H^2 \biggl(~
     \ln {gH\over \mu^2}~-~{1\over 2}~\biggr)
     ~+i~{3 \over 50 \pi}~g^2 H^2 ~+~\ldots \, ,
\end{equation}
and in 3 dimension (see \cite{noi,woloshyn,singer}), we have:
\begin{equation}
\label{chisara}
     V(H)~=~{1\over 2}H^2~-~{2-\sqrt{2}\over 3\pi^2}~
       (gH)^{3/2}
     ~-i~ {5\over 6\pi}~(gH)^{3/2}~+~\ldots \, .
\end{equation}

\section {INSTABILITY}

When for some field configuration there is an imaginary part
in the effective action this configuration is called unstable.
In this case
Unfortunately this is what happens for
non-Abelian gauge theories where the expansion
is around a saddle point in the functional space.
The second derivative operator of the action
has some negative eigenvalues and, when loop-expanded, it shows a
complex effective potential.
Therefore, the  validity of the loop expansion in these cases has
been questioned by Maiani et al \cite{maiani}.
These authors strongly argued that the
calculation of the effective potential in the
presence of a background field is actually
non-perturbative. When the background is not realizing a true
minimum there is no possibility for the existence of any
reliable perturbative expansion.
So far the only known non-perturbative
technique in field theory that can be made systematic
is the lattice approach.
More recently several works with the lattice approach
have appeared in the literature, both in four dimensions
\cite{ambjorn} and three dimensions \cite{woloshyn}.
Due to various technical reasons, mainly related to
the fact that the Savvidy ansatz in Eq.(\ref{abel})
is non-uniform, all these lattice works have yet
to yield a conclusive result.

Because we  expand around
a field configuration $\phi_c$ which is a saddle point
in the functional space, the second derivative operator of the action
will have at least one negative eigenvalue.
We use this operator to choose a basis in the vicinity of
$\phi_c$ that has as components the eigenvectors of this
operator. Let $\phi^s$ and $\phi^u$ designate the
stable and the unstable modes; these correspond, respectively, to the
positive and negative eigenvalues.
Hence, a configuration $\phi$ in the vicinity of
$\phi_c$ can be written as:
\begin{equation}
\label{filomena}
    \phi~=~\phi_c~+~\sum_i~c_i^s~\phi_i^s~+
    ~\sum_i~c_j^u~\phi_j^u
\end{equation}
where the index $i$ runs over the set of stable modes $\{ s \}$,
and  $j$ runs over the set of unstable modes $\{ u \}$.

We can hope that after splitting the
stable from the unstable modes, we will be able to perform a
perturbative expansion in the stable sector and handle
non-perturbatively the unstable one.
Due to the orthogonality of the stable and unstable modes,
the measure is simply defined now as
$ (d\phi)=(d\phi^s)~(d\phi^u)$,
and we can write $W[J]$ as:
\begin{eqnarray}
\label{guglielmo}
    e^{{i\over \hbar}W[J]}~&=~\int~(d\phi)~e^{{i\over \hbar}
    \{ S[\phi]+(J,\phi)\} }~= \nonumber \\
    &=~\int~(d\phi^s)(d\phi^u)~e^{{i\over \hbar}
    \{ S[\phi_c+\phi^u+\phi^s]+(J,\phi_c+\phi^u+\phi^s)\} }~=
    \nonumber \\
    &=~\int~(d\phi^u)~e^{{i\over \hbar}
    \{ S_{eff}[\phi_c+\phi^u]+(J,\phi_c+\phi^u) \} }  \, ,
\end{eqnarray}
where we have supposed that we were able to handle the
integration over the stable modes sector.

As carefully discussed  in reference \cite{maiani},
this is not a correct way to proceed.
It is true that
we were formally able to write the connected Green's functions
generating functional $W[J]$.
However, the Legendre transformation is now different.
In fact, now we are in presence of
many solutions, and we must
take the Legendre transform, choosing the $J$ that minimizes
$W[J]-(J,\phi)$ at fixed $\phi=\phi_c$.
Therefore, the effective potential is defined as:
\begin{equation}
\label{leggiadra}
    \Gamma[\phi_c]~=~\min_{ \{ J \} }~
    \biggl[ W[J]-(J,\phi) \biggr]
\end{equation}
This definition reduces to the usual expression for stable
configurations, but the
choices of $J$ may drive the system into a non-perturbative regime for
unstable $\phi_c$.

In the classical limit $\hbar \rightarrow 0$,
the integral
is still dominated by the saddle point
configuration $\phi_0$.
Now, however, the expression above
is no longer valid since there are no
restoring forces for the unstable mode.
What remains in equation is an effective action
\begin{equation}
\label{urania}
     S_{eff}[\phi_c+\phi^u]=S[\phi_c+\phi^u] + O(\hbar)
\end{equation}
which comprises the
effects of small fluctuations around $\phi_c+\phi^u$.
This is not what we were doing in the previous section when
we were took into account the small fluctuations around $\phi_0$
looking for
\begin{equation}
\label{plutonia}
      S_{eff}=S[\phi_c+\phi^u+\phi^s] + O(\hbar)
\end{equation}
Thus, we can conclude, following \cite{maiani}, that the problems are
that, one,
$\phi_0$ will definitely not dominate $W[J]$ and, two,
the small fluctuations around $\phi_0$ and $\phi_c+\phi^u$
have nothing to do with each other.

A consequence of this is that the Legendre transform, evaluated at
$\phi_c$, deviates from the minimum
by a dangerous term which is $O(\phi^u)$.
Hence, we no longer have the possibility of using
$\hbar$ as a parameter to expand the effective potential at $\phi_0$.
The non-perturbative effects become dominant and,
as a result, non-perturbative techniques,
such as the lattice regularization, must be used to find what replaces
the one-loop analysis which, by definition,  was obtained
by the help of the perturbation expansion.

If this argument is correct, then the Savvidy solution,
which was obtained
perturbatively, no longer provides a good picture.
Only a non-perturbative approach is reliable.

On the other hand, if the arguments of references
\cite{copenhagen}
are correct, the Savvidy solution may be a good approximation,
in the sense that
even if quantum fluctuations alter the vacuum from
the original chromomagnetic
configuration, the transition might not be too drastic.
If the imaginary part is negligible, in first approximation,
the vacuum can be thought
as microscopic domains of constant chromomagnetic
fields. This is analogous to the case of ferromagnetic
domains of iron below the Curie
temperature.
To be more precise, we cannot
prove the existence of domains without
studying the balance of the energy contributions of the walls
and of the domain bulks.
Nevertheless, from the fact that in the effective potential
a domain scale of constant chromomagnetic field exists
we might expect the existence of such domains.
Hence, Monte Carlo simulations will be able to
detect $gH_{min}$.
However, if the fluctuations do
occur very drastically, then the configurations will fluctuate so
violently that it will be very difficult to collect evidence
of a minimum.

A hybrid scenario might also occur.
For intense fields, we might have a non-trivial vacuum, and for weak
fields, the trivial vacuum, dominates.
This hybrid scenario
is the most difficult to investigate, because
it is not clear quantitatively when the strong field regime
is reached.
Moreover, for lattice simulations, this is the region
in which the samples are most difficult to collect.
At the present stage of
lattice simulations \cite{ambjorn,woloshyn} we can not exclude
consistence with this hybrid scenario, and it will be
extremely important to
to clarify this possibility with further analysis.

Note, however, that the existence of the imaginary part is
also related to the asymptotic freedom of the pure gauge theory.
In fact, in the absence of the
imaginary part, the ultraviolet limit of
the Savvidy solution implies that the beta function will be the same as
the one of a scalar particle of mass $m^2=2gH$.
It is just the imaginary part which prevents us after regulation
from rotating the integration contour in the complex plane
and thereby spoiling the asymptotic freedom.

\section {Comparison with Coleman and Weinberg Potentials }

In the discussion of the radiatively induced symmetry breaking,
\cite{coleman},
the question of the
trustworthiness of the non-trivial vacua is also addressed.
In particular, Coleman and Weinberg found
a non-trivial minimum in the one-loop effective potential of
the $\lambda \phi^4$ theory at
\begin{equation}
\label{sidney}
     \lambda ~\log~{\phi_c^2 \over M^2}~=~-{32 \over 3} \pi^2
     ~+~O(\lambda)
\end{equation}
but they noted immediately that the effective potential
is not calculable at this minimum in the loop expansion.
In fact, higher loop contributions bring higher powers of
$\lambda \log {\phi_c^2 \over M^2}$ and then,
independently of how small the coupling constant is,
the new minimum will lie outside the range of
validity of the one-loop approximation.

In the same article, the case of the scalar electrodynamics
theory was also addressed.
In this case, the effective potential is found to be
\begin{equation}
\label{erick}
     V~= {3~e^4\over 64\pi^2}~\phi_c^4~\log \biggl(~
     {\phi_c^2 \over <\phi >^2}~-{1\over 2}~\biggr)
\end{equation}
In this case the non-trivial vacuum must be trusted,
the loop expansion can be trusted at the minimum for the appropriate choice
of the coupling constants because the higher order terms in $\lambda $
can be balanced by the contributions of the order
$e^4\log~{\phi_c \over M}$.

However, there is a fundamental difference between
these cases and the non-Abelian gauge theories.
For the above cases, the effective potential can be evaluated
directly because there is no instability,
while the opposite is true of
the non-Abelian gauge theory.
Consider for example the $\lambda \phi^4$ theory in the broken
phase ($\mu^2 < 0$).
A possible source of instability comes from
the imaginary part of the contribution of the effective potential,
given by:
\begin{equation}
\label{ester}
     \int ~dk~\log \bigl( k^2~+~V''(\phi_c) ~\bigr) ~=~
     \int ~dk~\log \bigl( k^2~+~\mu^2~+{\lambda \over 2}
     \phi^2_c ~\bigr)
\end{equation}
when the logarithm is integrated for small $k$.
However, because $\phi_c=\phi_c[J]$ it is always possible to find a
set of values of $J$ sufficiently large
for which there is a region where
${\lambda \over 2} \phi^2_c > \mu^2~$
and then it is possible to perform the
integration without the imaginary part.
Obviously, the existence of this region is not enough:
we must reach the limit $J \rightarrow 0$ while
remaining in the stable region.
This is possible here by
reaching the minimum via
$\phi_c[J_1] > \phi_c[J_2<J_1] > \phi_c[J=0] $,
while remaining away from the unstable region, that is
$\phi < {1\over 2}\phi_c[J=0]$.
The lesson of this simple observation is that
it is the quartic part, $\phi^4$, of the effective potential
which stabilizes the vacuum.

In non-Abelian gauge theories the situation is different.
The same analysis might be performed with the vocabulary:
\begin{eqnarray}
\label{sfracassa}
    -~\mu^2~& \longleftrightarrow ~2gH \\
    {\lambda \over 2} \phi^2_c~& \longleftrightarrow ~6g A^2_c
\end{eqnarray}
The first identification comes by inspection of the
ultraviolet part of the effective potential,
or equivalently, by neglecting the
imaginary part and comparing the beta function with the one
of the scalar theory.
The second correspondence comes directly from the quartic part of the
action.

For the non-Abelian theory the equivalent contribution of
(\ref{ester}) is:
\begin{equation}
\label{ester2}
     \log \bigl(~k^2~-~gH~ \bigr)
\end{equation}
and not the ``safe''
\begin{equation}
\label{eletta}
          \log \bigl(~k^2~-~gH~+~6g A_c^2~ \bigr)
\end{equation}
because the last term is not gauge invariant.
Hence, there is no region with stable minimum, as opposed to the
scalar field theory. So unstable fluctuations must always be present.

This also teaches us that variational methods will not work for
the vacuum properties of non-Abelian gauge theories.
The variational method is very useful and powerful
in the case of stable configurations.
However, in my opinion, to use the variational method
to analyze situations where unstable
configurations are dominant is extremely hazardous.

The next natural question is if higher order contributions
stabilize the instability.
Unfortunately, we  must perform a partial resummation of
the loop expansion to clarify this point for non-Abelian models.
At the present there is no evidence for a window
for $J$ where the argument of the logarithm could be made positive.

Consider now the case of SU(2) coupled to a fermion.
In this case,
there is a contribution to
the one-loop effective potential
from the graph with four gluonic
legs connected by a fermion loop.
The contribution of this diagram is analogous
to the one in the Abelian case (see \cite{landau}),
and for low momentum transfer,
will give an extra term in the effective potential
equal to
${8\over 135 } {(qe)^4\over m^4} H^4 $,
where $m$ is the mass,
and $(qe)$ is the charge of the fermion.
The difference in the coefficient of $H^4$
with respect to the Abelian
case is due to the fact that for SU(2) there is an
extra factor $f^{bcd}$ at each vertex.
Hence, for a sufficiently small, but non zero $m$,
there is a region of $J$ where the instability is
not present.
The issue in this case is if it is possible
to reach $J=0$ remaining in a stable region.
There is a physical argument that might suggest that this is
possible: adding gauge bosons at the same space-time point
will strongly polarize the vacuum and eventually generate
a chromomagnetic domain, but in the presence of fermions,
the vacuum will
soon create fermion anti-fermion pairs at the expense of the
energy of the gluons.
Depending on which of these two phenomena
occurs first the vacuum structure will be different.
It will be interesting to explore this situation in more detail.
Unfortunately the realistic lattice simulations to investigate
the vacuum structure with quarks are unfeasible by the present techniques.

\section{CONCLUSIONS}

To summarize, we have analyzed the effective potential for
a non-Abelian background field in SU(2) Yang-Mills theory.
The result is found to be
qualitatively similar to the Savvidy ansatz,
both in real part, which indicates a
spontaneous generation of the color magnetic field, and imaginary
part, which signals the instability of the background field
as vacuum configuration under the loop expansion.
Given the qualitative
similarity between the three dimensional and four dimensional
effective potential in the Savvidy ansatz, we conclude that the
effective potential is insensitive to the coordinate singularity,
${\rm Det}B_i^a=0$, if it indeed exists.

Rather, we realize that the presence of the imaginary part
in non-Abelian theories is due to the
incompatibility in the action between gauge invariance
and stabilizing terms.

Technical questions related
to the linear term and the gauge choice were illustrated,
and in particular, we show the dependence of the
one-loop effective potential of
the gauge fixing parameter $\alpha$ when the loop expansion
is performed around a non trivial background.

It is important to recognize that the effective potential
for these kinds of background
would be a well-defined problem if the functional
were evaluated non-perturbatively. The appearance of the imaginary
part in the effective potential is only caused by the loop expansion.
In other words, when we were doing steepest descent approximation,
we were expanding at a saddle point. In addition, even if the
expansion point were a true minimum, the stationary solution of the
effective potential calculated up to a finite loop order could not
be trusted quantitatively, due to the fact that at the stationary
point of the effective potential the higher order terms become as
important as lower order terms and hence the loop expansion breaks
down. The final answer has to be settled by a non-perturbative
means, such as the lattice simulation as mentioned in the Introduction.

In fact, a lattice generalization of the background field method
is rather straightforward. Let us consider
\begin{equation}
     Z_L[J_\mu(x),B_\mu(x)]\equiv\int dU_\mu(x)\,
     \exp\biggl\{-S[U_\mu(x)B_\mu(x)]+
      {\rm Tr}J_\mu(x) f[U_\mu(x)]\biggr\}\, ,
\end{equation}
where $U_\mu(x)$ is the standard link variable, $B_\mu(x)$ is the
background link variable, $S$ is the usual Wilson lattice action,
$J_\mu(x)$ is a matrix valued external current, and $f[U]$ is an
arbitrary function satisfying $gf[U]g^\dagger=f[gUg^\dagger]$ for
any unitary matrix $g$. Using the property of the invariance
under an unitary transformation for the Haar measure, one can easily
verify the following:
\begin{equation}
        Z_L[\tilde{J}_\mu(x),\tilde{U}_\mu(x)]
       \bigg|_{\tilde{U}_\mu(x)=g(x)U_\mu(x)g^\dagger(x+\mu),\,
        \tilde{J}_\mu(x)=g(x)J_\mu(x)g^\dagger(x)}
        =Z_L[J_\mu(x),U_\mu(x)]\, ,
\end{equation}
a lattice version of Eq.(\ref{ginv}). A Legendre transform of
$Z_L[J_\mu,U_\mu]$ would lead to a gauge invariant effective action,
provided that the induced gauge field is constrained to have zero
expectation value by adjusting $J_\mu(x)$, just as in the continuum
case. The remarkable thing here is that we do not need to fix the
gauge on a lattice and therefore the resulted effective action is
unique for a given choice of the functional form of $f$.

As mentioned above the non-Abelian background field ansatz
can be conveniently realized on a lattice. The constant
nature of the ansatz avoids problems with the boundary condition
and non-uniformness of the lattice constant effect due to the
linear rising ansatz of Savvidy.
The hope is that intensive
lattice simulations with the non-Abelian background
will clarify the picture.
Since the non-Abelian background only involves
one parameter $h$ it may not be difficult to find a way to adjust the
external current $J_\mu$ to ensure a vanishing of the expectation
value for the induced quantum field.

\acknowledgements
This work was possible thanks to the collaboration of
Suzhou~Huang, Ken~Johnson and Janos~Polonyi
to whom I am very grateful.
It is my pleasure to thank Kerson Huang,
H.~B.~Nielsen, H.~Trottier, P.~van Baal,
for many very fruitful discussions.

\end{document}